\documentclass[aps,twocolumn,prl,superscriptaddress,floatfix,showpacs,tightenlines]{revtex4}
\usepackage{amssymb}
\usepackage{amsmath}
\usepackage{graphicx}
\usepackage{epsfig}

\begin{document}

\title{Macroscopic Quantum Criticality in a Circuit QED}
\author{Y.D. Wang}
\affiliation{Frontier Research System, The Institute of Physical
and Chemical Research (RIKEN), Wako-shi, Saitama 351-0198, Japan }
\affiliation{Institute of Theoretical Physics, Chinese Academy of
Sciences, Beijing, 100080, China}
\author{H.T. Quan}
\affiliation{Frontier Research System, The Institute of Physical
and Chemical Research (RIKEN), Wako-shi, Saitama 351-0198, Japan }
\affiliation{Institute of Theoretical Physics, Chinese Academy of
Sciences, Beijing, 100080, China}
\author{Yu-xi Liu}
\affiliation{Frontier Research System, The Institute of Physical
and Chemical Research (RIKEN), Wako-shi, Saitama 351-0198, Japan }
\author{C.P. Sun}
\affiliation{Frontier Research System, The Institute of Physical
and Chemical Research (RIKEN), Wako-shi, Saitama 351-0198, Japan }
\affiliation{Institute of Theoretical Physics, Chinese Academy of
Sciences, Beijing, 100080, China}
\author{Franco Nori}
\affiliation{Frontier Research System, The Institute of Physical
and Chemical Research (RIKEN), Wako-shi, Saitama 351-0198, Japan }
\affiliation{Center for Theoretical Physics, Physics Department,
CSCS, The University of Michigan, Ann Arbor, Michigan 48109-1040}
\date{\today}

\begin{abstract}
Cavity quantum electrodynamic (QED) is studied for two
strongly-coupled charge qubits interacting with a single-mode
quantized field, which is provided by a on-chip transmission line
resonator. We analyze the dressed state structure of this
superconducting circuit QED system and the selection rules of
electromagnetic-induced transitions between any two of these
dressed states. Its macroscopic quantum criticality, in the form
of ground state level crossing, is also analyzed, resulting from
competition between the Ising-type inter-qubit coupling and the
controllable on-site potentials.
\end{abstract}

 \maketitle

\emph{Introduction.---} In cavity quantum electrodynamics
(QED)~\cite{cqed}, the coupling effects of two atoms~\cite{2cqed}
inside a cavity have been theoretically studied to explore exotic
quantum coherent phenomena~\cite{scully}, e.g., coherent
population trapping and  dark states. However, the weak
dipole-dipole interaction between atoms make it difficult to
experimentally demonstrate these phenomena.

Recent experiments using on-chip superconducting
qubits~\cite{phy,NEC,2flux} show the strong-coupling effects
between: i) two superconducting qubits~\cite{NEC,2flux}; ii) a
charge qubit and a superconducting transmission line resonator
(TLR)~\cite{you1,Yale};  and iii) a flux qubit and an LC
oscillator~\cite{Ljjq,wang}. The later two ones have been referred
to as circuit QED. Moreover, recent experiments~\cite{Yale}
demonstrated Rabi splitting and AC Stark shift in these QED
circuits. Combining these approaches~\cite{NEC,2flux,you1,Yale},
we propose how to observe the above mentioned quantum coherence
effects using two coupled artificial atoms, inside a cavity,
instead of two natural atoms~\cite{2cqed}.

Here, we present a circuit QED architecture for two
capacitively-coupled charge qubits, interacting with a single-mode
quantized field. We not only study the influence of strong
inter-qubit coupling on quantum coherent effects, but also explore
its macroscopic quantum criticality, via level-crossing,  for
quantum phase transition (QPT)~\cite{qpt}. Here, the
non-analyticity of the ground state is depicted by the fact that
the eigenstates are independent of the inter-qubit coupling, while
the corresponding eigenvalues depend on the inter-qubit coupling.
Of course, a rigorous QPT can only be realized in the
thermodynamic limit, but some of the basic features of QPT can
still be demonstrated in the form of level crossings for a system
of few qubits~\cite{qpt,qian}. A toy-system for QPT, with two
qubits, has been experimentally studied using NMR~\cite{du}.

\begin{figure}
\includegraphics[bb=59 494 455 729, width=8 cm, clip]{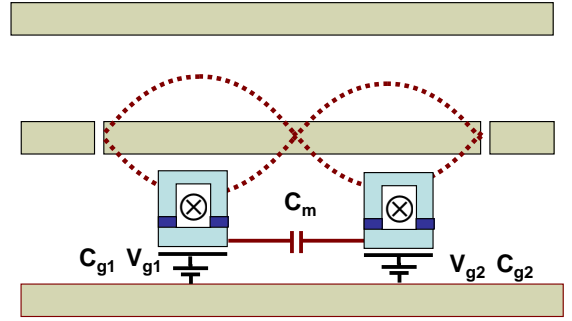}
\caption{(color online). Schematic setup of a circuit QED with two
capacitively-coupled charge qubits, interacting with a
transmission line resonator. The charge qubits are placed between
superconducting lines and located at the antinodes of a
single-mode magnetic field. The couplings between two qubits and a
single-mode quantized field are realized by the magnetic field
through the qubit loops. } \label{fig1}
\end{figure}

In contrast with a previous investigation for QPT~\cite{qpt}, the
two Ising-type-coupled charge qubits in our macroscopic quantum
system are also coupled to a TLR. Thus,  for the first time, we
incorporate the dressed-state structure in the QPTs. Here, we
consider a minimal circuit QED model to simulate the quantum
critical behavior of the ground state.

\emph{Circuit QED model with two qubits.---}  Our circuit QED
system is schematically shown in Fig.~\ref{fig1}. Two identical
SQUID-based charge qubits are capacitively coupled to each other
with the coupling strength $J= e^{2}C_{m}/\left( 2C_{\Sigma
}^{2}-2C_{m}^{2}\right)$ through a capacitance $C_{m}$, where
$C_{\Sigma }$ is the sum of the capacitances connected to a single
Cooper pair box. The level-spacing of each qubit is $\omega
_{a}=2E_{c}\left( C_{g}V_{g}-1/2\right)$, with capacitance
$C_{g}$, bias voltage $V_{g}$, and $E_{c}=2e^{2}C_{\Sigma }/\left(
C_{\Sigma }^{2}-C_{m}^{2}\right) $. The two qubits are coupled to
a single-mode quantized field which is realized as a 1D TLR with
resonant frequency $\omega =n_{0}\pi /(L\sqrt{lc})$. Here $n_{0}$
and $L$ are the mode number of the resonant mode and the length of
the TLR; $l$ and $c$ are the inductance and capacitance per unit
length of the TLR. When the two dc SQUIDs are placed at
$x_{n}=nL/n_{0}$, the coupling between the qubit and TLR is
induced by the quantized magnetic field threading the dc SQUID.
The model Hamiltonian $H=H_{Q}+H_{C}$ includes the Ising part
\begin{equation}
H_{Q}=\frac{1}{2}\omega _{a}\left( \sigma _{z}^{\left( 1\right) }+\sigma
_{z}^{\left( 2\right) }\right) +J\sigma _{z}^{\left( 1\right) }\sigma
_{z}^{\left( 2\right) }
\end{equation}
and the Jaynes-Cummings terms
\begin{equation}
H_{C}=\omega a^{\dag }a+\frac{g}{\sqrt{2}}\left[\left(\sigma
_{+}^{(1) }+\sigma _{+}^{(2) }\right)a+\mathrm{H.c.}\right].
\end{equation}
Hereafter, $\hbar=1$,  the Pauli matrices $\sigma _{z}=|\!\uparrow
\rangle \langle\uparrow \!| -|\!\downarrow \rangle
\langle\downarrow\!| $, $\sigma _{+}=|\!\uparrow\rangle\langle
\downarrow\!|$ and $\sigma _{-}=|\!\downarrow\rangle
\langle\uparrow\!|$ are defined by the charge eigenstates
$|\!\uparrow \rangle $ and $|\!\downarrow \rangle $ denoting $1$
or $0 $ excess Cooper pair state, respectively. $a^{\dag}$ $(a)$
is the creation (annihilation) operator of the resonant mode. The
coupling strength is $g=SE_{J}\sqrt{\hbar l\omega }/(\Phi _{0}
d\sqrt{L})$ with the tunnelling energy $E_{J}$, the enclosed area
$S$ of the dc SQUID, the distance $d$ between the dc SQUID and the
transmission line, and the flux quantum $\Phi _{0}=\hbar /2e$.

The symmetry of two qubits enables us to rewrite the Hamiltonian
$H$ as a function of the total spin operators
$\mathbf{S}=(\overrightarrow{\sigma}^{(1)}
+\overrightarrow{\sigma}^{(2)})/2$ and $S_{\pm }=\sigma _{\pm
}^{(1)}+\sigma _{\pm}^{(2)}$, e.g., $
\sigma^{(1)}_{z}\sigma^{(2)}_{z}=2S_{z}^{2}-1$. The dynamical
symmetry SO(3) of the Hamiltonian $H$ results in a direct sum
decomposition of the two-qubit Hilbert space $V$, spanned by
$\{|\uparrow \uparrow \rangle ,\,|\downarrow \downarrow\rangle
,\,|\downarrow \uparrow\rangle,\,|\uparrow \downarrow \rangle \}$;
i.e., $V=V^{(0)}\oplus V^{(1)}$. Here, $V^{(0)}$ is spanned by the
singlet $|\psi ^{-}\rangle =(|\uparrow\downarrow\rangle
-|\downarrow \uparrow \rangle)/\sqrt{2}$, while $V^{(1)}$ is
spanned by the triplet $|\uparrow\uparrow\rangle$,
$|\psi^{+}\rangle=(|\downarrow\uparrow\rangle+|\uparrow
\downarrow\rangle)/\sqrt{2}$, and $|\downarrow \downarrow \rangle
$. $V^{(0)}$ and $V^{(1)}$ are invariant under $H$. Thus, the
total Hamiltonian $H$ can be decomposed into a quasi-diagonal
matrix with two blocks. Obviously, the symmetric couplings of the
two qubits to the quantized field do not induce transitions
between the singlet $|\psi ^{-}\rangle$ in $V^{(0)}$ and any other
states in the space $V^{(1)}$. This is because the collective spin
operator $S_{\pm }=\sigma _{\pm }^{(1)}+\sigma _{\pm}^{(2)}$ can
only change the state vectors within an irreducible subspace.
Therefore, here, we denote the singlet $|\psi^{-}\rangle$ as a
``dark state"~\cite{scully}. This consideration, based on group
representations, automatically predicts the coherent population
trapping in this superconducting macroscopic quantum system,
without resorting to any dynamical evolution calculations for the
natural atoms~\cite{2cqed}. The population on the singlet
$|\psi^{-}\rangle$ will be trapped to keep its initial value, due
to the coherent cancelling of the two transitions from both,
$|\uparrow \downarrow\rangle $ and $|\downarrow \uparrow\rangle $,
to any state.

\emph{Dressed Spectrum.---}For quantum criticality, the Ising-type
Hamiltonian $H_{Q}$ has been extensively studied in the
thermodynamic limit, i.e., the case with infinite
qubits~\cite{qpt}. To demonstrate this with a few qubits, we
consider the eigenvalues of $H_{Q}$: $E_{|\psi ^{+}\rangle }=-J$,
$E_{|\uparrow \uparrow \rangle }=J+\omega _{a}$, $E_{|\psi
-\rangle }=-J$ and $ E_{|\downarrow \downarrow\rangle }=J-\omega
_{a}$. We use the dimensionless parameter $\xi =\omega _{a}/J$  to
describe the QPT character of the ground state. If $\xi$ changes
from region $\xi <-2$ to $-2<\xi <2$, and then to $\xi
>2$,  the ground state of $H_{Q}$ changes from the state
$|\uparrow \uparrow\rangle $ to the state $|\psi^{+}\rangle $
$(|\psi^{-}\rangle)$ and then to $|\downarrow \downarrow\rangle $.
Correspondingly, the ``order" of the magnetic system changes from
ferromagnetic to antiferromagnetic through the maximal entangled
state $|\psi^{+}\rangle$ ($|\psi^{-}\rangle$) corresponding to the
point $\xi=0$.

For a weak perturbation, not commuting with $S_{z}$ or $H_{Q}$,
the ground state is still determined by the longitudinal field
controlled by the gate voltages. Now, we consider the effect of
the strong interaction on the ground state when the $H_{Q}$ system
interacts with a single-mode quantized field. The above
discussions show that the quantized field can only mix qubit
states, in the invariant subspace $V^{(0)}$ or $V^{(1)}$, to form
dressed states,  but it cannot induce the transitions from
$V^{(0)}$ to $V^{(1)}$. Therefore, in the qubit subspace
$V^{(0)}$,  the eigenstates $|\varphi_{n}^{(s)}\rangle$ of the
Hamiltonian $H$ are the product states of the singlet
$|\psi^{-}\rangle$ and photon number states $|n\rangle$, i.e.,
$|\varphi_{n}^{(s)}\rangle\equiv |n,\psi^{-}\rangle$, which
correspond to the eigenvalues $E_{n}^{(s) }=n\omega-J$, with
$n=0,\,1,\,2,\cdots$.

However, in the qubit subspace $V^{(1)}$, when we consider the
state mixture induced by the quantized field, the Hamiltonian $H$
possesses the following invariant subspaces
$W^{(0)}:\{|0,-1\rangle \equiv |0,\downarrow \downarrow\rangle
\}$, $W^{(1) }:\{ |0,0\rangle \equiv |0,\psi
^{+}\rangle,\,|1,-1\rangle \equiv|1,\downarrow \downarrow
\rangle\} $, $W^{(n+1)}:\{|n-1,1\rangle \equiv |n-1\rangle \otimes
|\uparrow \uparrow\rangle $,\ $|n,0\rangle \equiv |n\rangle
\otimes |\psi ^{+}\rangle $, $|n+1,-1\rangle \equiv |n+1\rangle
\otimes |\downarrow \downarrow\rangle \}$ for $n=1,2,3,\cdots $.
Then, the Hamiltonian $H$ can be diagonalized in each
quasi-diagonal block $W^{(n)}$. The 1-D block $W^{(0)}$ only
contains the eigenstate $|\varphi_{0}^{(0)}\rangle =|0,-1\rangle $
with eigenvalue $E_{0}^{(0)}=J-\omega _{a}$. The 2-D block
$W^{(1)}$ is spanned by two eigenstates
\begin{eqnarray}
\left|\varphi _{0}^{(+)}\right\rangle &=&\cos(\theta/2)|0,0\rangle
-\sin (\theta/2)|1,-1\rangle, \notag \\
\left|\varphi _{0}^{(-)}\right\rangle &=&\sin (\theta /2)|
0,0\rangle +\cos (\theta/2)|1,-1\rangle
\end{eqnarray}
with eigenvalues $E_{0}^{(\pm )}=\pm \sqrt{J^{2}+g^{2}}$. The
eigenvalues of the 3D blocks $W^{( n+1) }(n\geqslant 1)$ can be
explicitly solved in the resonant case $\omega=\omega _{a}$. In
this condition, the eigenvalues are $ E_{n}^{(0)}=n\omega +J$ and
$E_{n}^{(\pm )}=n\omega \pm N_{n}(g)$, which correspond to the
eigenstates
\begin{eqnarray}
\left\vert \varphi _{n}^{\left( 0\right) }\right\rangle
&=&\sqrt{\frac{n+1}{ 2n+1}}\left\vert n-1,1\right\rangle
-\sqrt{\frac{n}{2n+1}}\left\vert
n+1,-1\right\rangle ,  \notag \\
\left\vert \varphi _{n}^{\left( \pm \right) }\right\rangle
&=&\sqrt{\frac{ ng^{2}}{2\Omega _{n\pm }\left( g\right)
}}\left\vert n-1,1\right\rangle \pm
\sqrt{\frac{N_{n}\left( g\right) \mp J}{2N_{n}\left( g\right) }}
\left\vert n,0\right\rangle  \notag \\
&&+\sqrt{\frac{(n+1)g^{2}}{2\Omega _{n\pm }\left( g\right)
}}\left\vert n+1,-1\right\rangle,
\end{eqnarray}
where $N_{n}\left( g\right) =\sqrt{J^{2}+\left( 2n+1\right)
g^{2}}$ and $\Omega _{\pm }\left( g\right) =N_{n}^{2}\left(
g\right) \mp JN_{n}\left(g\right)$.

\begin{figure}
\includegraphics[bb=20 182 552 570, width=7.5 cm, clip]{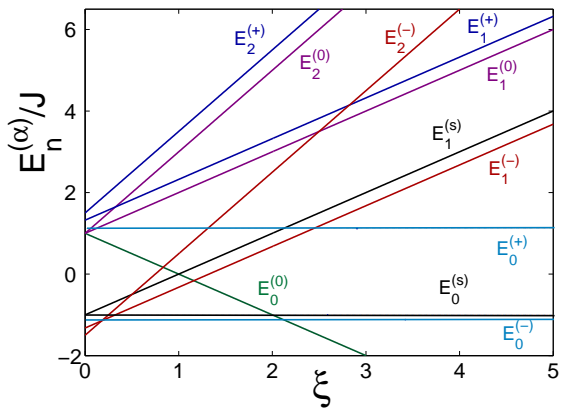}
\caption{(color online). Rescaled eigenvalues
$\{E_{n}^{(\alpha)}/J\}$ for $\alpha =0,\,\pm,\,s;\,n=0,\,1,\,2$
of dressed ``atom"-photon states versus the rescaled longitudinal
field $\xi=\omega/J$, in the resonant case $\omega=\omega_{a}$ and
for a parameter ratio $g/J=0.5$.}\label{fig2}
\end{figure}

\emph{Quantum criticality due to level crossings.---}The
eigenvalues $\{E_{n}^{(\alpha)}\}$ with $\alpha
=0,\,\pm,\,s;\,n=0,\,1,\,2,\,\cdots$ form a complete set for the
energy spectrum. Fig.~\ref{fig2} shows a few lowest eigenvalues
versus the effective longitudinal field $\xi =\omega/J$, when
$\omega_{a}=\omega$ and $g/J=0.5$. However, for very weak coupling
constant $g$ (e.g., $g/J=0.05$), some near-neighbor eigenvalues
are almost degenerate, for instance,
$E_{1}^{(+)}=\omega+J\sqrt{1+3(g/J)^2}\approx\omega+J=E_{1}^{(0)}$,
$E_{2}^{(0)}\approx E_{2}^{(+)}$, and so on. Comparing with the
case without the coupling to the quantized field, i.e., $g=0$, the
distribution of the spectral structure of the dressed two-qubit
system becomes very complicated. Usually, the cavity field
coupling $g$ can be very small with respect to $J$ and $\omega$.
In this case, $E_{n}^{(\pm)}$ can be approximated by
\begin{equation}\label{eq:7}
E_{n}^{(\pm)}=n\omega \pm \lbrack J+\delta (n)].
\end{equation}
Eq.~(\ref{eq:7}) predicts that level shifts $\delta (n)=(
n+1/2)g^{2}/2J$ can be enhanced by the photon number of the
quantized field.

\begin{figure}
\includegraphics[bb=12 182 590 572, width=8.5 cm, clip]{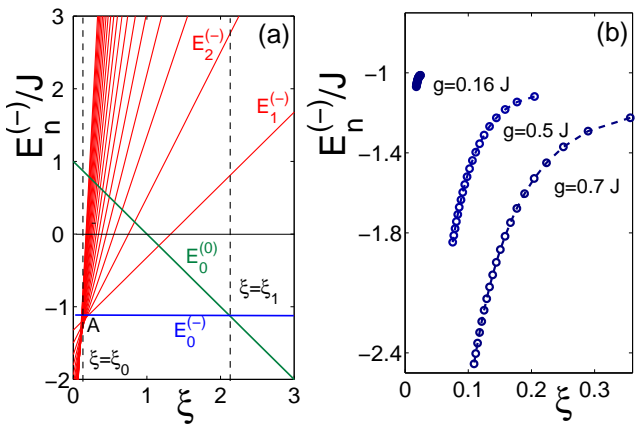}
\caption{(color online). Quantum criticality described by
level-crossing points. (a) The rescaled $20$ energy levels
$E_{n}^{(-)}/J$  ($n=0,\, 1,\cdots,\,19$) versus $\xi=\omega/J$
with $g/J=0.5$. The blue line denotes
$E_{0}^{(-)}/J=\sqrt{1-(g/J)^2}$; red lines show $E_{n}^{(-)}/J$,
with $n\neq 0$. They converge to a critical point $A$. As a
reference,  the green line is for $E_{0}^{(0)}/J=1-(\omega/J)$.
(b) The same energy levels $E_{n}^{(-)}/J$ as in (a) versus the
crossing points $\xi=\omega/J$ for near-neighbor levels with
different couplings $(g=0.7 J,\, 0.5 J,\, 0.16 J)$ of qubits to
the cavity field. }\label{fig3}
\end{figure}

To study quantum criticality, we now analyze the distribution of
the level-crossing points. We plot the rescaled $20$ lowest
eigenvalues $E_{n}^{(-)}/J$ in Fig.~\ref{fig3}(a) for
$n=0,\,1,\cdots,\, 19$. It can be seen that all the energy levels
approximately meet together at a certain critical point $A$, which
is an intrinsic quantum critical point. This is because in the
weak-cavity-field coupling limit, i.e., $g\ll J, \,\omega$, all
energy levels $E_{n}^{(-)}$ ($n=0,\,1,\,\cdots$) linearly depend
on $n$, and thus they are approximately degenerate at a fixed
point $\xi _{0}=g^{2}/J^{2}$, independent of $n$. This feature is
further demonstrated in Fig.~\ref{fig3}(b). We plot the
eigenvalues of the level crossing points for each pair of
near-neighbor $E_{n}^{(-)}$ and $E_{n+1}^{(-)}$
($n=0,\,\cdots,\,19$) for different transverse coupling constants
$g$. Fig.~\ref{fig3}(b) shows that the crossing points of
different levels will slightly change along a certain curve for a
large coupling constant $g$. However, when $g$ gets smaller, the
distribution of those crossing points gets more compact. In the
limit of very weak cavity-field-coupling, those crossing points
converge to the fixed point $A$ with $\xi_{0}=g^2/J^2$.

Therefore, when $g\ll J, \,\omega$, the energy spectrum can be
divided into three parts (shown in Fig.~\ref{fig3}(b)): (I) $0<\xi
\leq \xi _{0}$; (II) $\xi _{0}<\xi \leq \xi _{1}$; and (III) $\xi
>\xi _{1}$, where $\xi _{1}=1+\sqrt{1+\xi _{0}}$ is a degenerate
point of $E_{0}^{(0)}$ and $E_{0}^{(-)}$. In the regions of (II)
and (III), the system has definite ground states
$|G^{(\text{II})}\rangle =|\varphi _{0}^{(-)}\rangle $ and
$|G^{(\text{III})}\rangle =|\varphi_{0}^{(0)}\rangle$,
respectively. It is very exotic that the system does not have a
ground state in the region (I). This is because $E_{n}^{(-)}$ has
no lower bound in the region (I). Thus, the critical point
$\xi=\xi _{0}$ is an intrinsic singular point, which is different
from the generic critical point $\xi _{1}$. The existence of this
intrinsic singular point $\xi_{0}$ can be verified via the
spectrum, generated by the transitions from excited states to the
ground state. As seen from Fig.~\ref{fig3}(a), the spectrum is
almost continuous when $\xi$ approaches $\xi_{0}$ from the side of
$\xi>\xi _{0}$. The discrete spectrum appears when $\xi$ is far
away from $\xi _{0}$. This feature serves as an {\it experimental
way to detect the intrinsic singular point} in the weak coupling
limit, as mentioned above. Though we concentrate on the resonant
case in the above discussions, the results can be generalized to
the off-resonant case.

\emph{Level transitions and widths of spectral lines.--- } Above,
we showed that quantum critical phenomenon can be observed from
the transitions between different dressed states. Now, we further
study the optical selection rules for the above different dressed
energy levels. We consider that the dressed system interacts with
a multi-mode bath, then the interaction Hamiltonian $H_{p}$
between the dressed system and bath is
\begin{equation}
H_{p}=\sum_{k}\left(g_{k}^{(1)}\sigma _{+}^{(1)}+g_{k}^{(2)
}\sigma _{+}^{(2)}\right) a_{k}+{\rm H.c.}.
\end{equation}
The free Hamiltonian $H_{b}$ of the multi-mode bath is
$H_{b}=\sum_{k}\hbar \omega _{k}a_{k}^{\dag}a_{k}$. $a_{k}^{\dag
}$ ($a_{k}$) are the creation (annihilation) operators of the
$k$th bath mode with the angular frequency $\omega_{k}$. The bath
can be the quasi-normal modes of the TLR or the multi-mode
external electromagnetic field.

When $g_{k}^{(1) }=g_{k}^{(2)}$, $H_{p}$ cannot induce the
transitions between two different diagonal blocks $V^{(0)}$ and
$V^{(1)}$ of the Hamiltonian $H$; it can only induce transitions
between different states in the same block. However, when
$g_{k}^{(1)}\neq g_{k}^{(2)}$, the perturbation $H_{p}$ will break
the original invariant subspaces $ V^{(0)}$ and $V^{(1)}$.
Transitions between the different subspaces $V^{(0)}$ and
$V^{(1)}$ are possible. This leads to the mixture of the different
diagonal blocks of the Hamiltonian $H$ discussed above.

As known in conventional cavity QED, many typical strong-coupling
phenomena, e.g., the Rabi splitting, actually refer to transitions
between the different invariant subspaces. The corresponding width
of the spectral line can be determined by the matrix elements
$\langle \varphi _{n}^{(\alpha) }| H_{p}|\varphi
_{m}^{(\beta)}\rangle$ through the Fermi golden rule. We can
analyze this problem in detail in different quantum critical
regions. For example, in region (III), $|\varphi_{0}^{(0) }\rangle
$ is the ground state, and $|\varphi _{0}^{(-)}\rangle $ and
$|\varphi _{1}^{(-)}\rangle $ are the first two excited states.
Transitions from $|\varphi _{0}^{(-)}\rangle $ and $|\varphi
_{0}^{(s) }\rangle $ to $|0,-1\rangle $ will lead to a double-peak
Rabi split. The distance between the center of the two peaks is
determined by the energy difference $\sqrt{J^{2}+g^{2}}-J$ between
two excited states. For instance, if we take $J\sim 4$
GHz~\cite{NEC} and $g\sim 2$ GHz~\cite{ying}, this Rabi splitting
is about $0.47$ GHz; however, for a small cavity field coupling
constant, e.g., $g\sim 0.2$ GHz, this Rabi splitting is about $5$
MHz.

We can also evaluate the ratios of the line widths, which are
determined by the damping rates. For example, the ratio of the
damping rates $\gamma _{1}$ (from $|\varphi _{0}^{(-) }\rangle $
to $|\varphi _{0}^{(0) }\rangle $) and $\gamma _{2}$ (from
($|\varphi _{0}^{(s) }\rangle $ to $|\varphi _{0}^{(0) }\rangle $)
is
\begin{equation}
\frac{\gamma _{1}(\omega _{1}) }{\gamma _{2}(\omega _{2}) }=\left[
\sin \left(\frac{\theta }{2}\right)\frac{G^{+}(\omega _{1}) \rho
_{1}(\omega _{1}) }{G^{-}(\omega _{2}) \rho _{2}(\omega _{2})
}\right] ^{2},
\end{equation}
where $\rho _{l}(\omega) $ is the given spectral density of the
bath, $\omega _{l}=\omega -(2-l)(J+\sqrt{J^{2}+g^{2}})$ ($l=1,2$)
and $G^{\pm }(\omega_{k})=g_{k}^{(1) }\pm g_{k}^{(2) }$. The above
quantitative results can also be tested by future experiments.

\emph{Conclusions.---} We analyze the dynamical symmetry of two
strongly-coupled charge qubits, interacting with a single-mode
quantized field. There exists a coherent trapped two-qubit singlet
state (``dark state"). This ``dark state" is not affected by any
other states when two qubits are symmetrically coupled to both the
bath and the quantized field. However, when the symmetry is
broken, the ``dark state" is no longer ``dark". The transitions
from this ``dark state" to other states become possible. By
analyzing the Rabi splitting, the level shift in the limit of
weak-cavity-field coupling, and the dressed state structure of the
transmission spectrum, we can probe the coherent coupling effect
between the two-qubit system and a single-mode quantized field.

We also study the influence of the strong inter-``atom" coupling
and atoms-to-quantized-field couplings on the quantum criticality
through level crossing.  Specially, an intrinsic singular point is
found in the limit of weak-cavity-field coupling. This point is
characterized by the discreteness of the spectra in some critical
regions. Our study can be easily generalized to the case of the
non-resonant interaction between the cavity field and two qubits.
We hope that our proposal can further motivate experiments on the
circuit QED with two strongly coupled qubits.

We acknowledge the partial support of the US NSA and ARDA under
AFOSR contract No. F49620-02-1-0334, and the NSF grant No.
EIA-0130383. The work of CPS is also partially supported by the
NSFC and FRP of China with No. 2001CB309310.

\end{document}